\documentclass[traditabstract]{aa}  
\usepackage{graphicx}
\usepackage{natbib}
\usepackage{txfonts}
\begin{document}
\title{Solar-like oscillations in the G8\,V star $\tau$~Ceti\thanks{Based
on observations collected at the European Southern Observatory, La Silla,
Chile (ESO Programme 74.D-0380)}}

   \subtitle{}

   \author{T.C.~Teixeira\inst{1,2} \and 
H.~Kjeldsen\inst{2} \and 
T.R. Bedding\inst{3} \and 
F.~Bouchy\inst{4} \and
J.~Christensen-Dalsgaard\inst{2} \and 
M.S.~Cunha\inst{1} \and  
T.~Dall\inst{5} \and
S.~Frandsen\inst{2} \and
C.~Karoff\inst{2} \and
M.J.P.F.G.~Monteiro\inst{1,6} \and 
F.P.~Pijpers\inst{2,7}          }

   \offprints{Tim Bedding}

   \institute{Centro de Astrof{\'\i}sica da Universidade do Porto, Rua das
Estrelas, 4150-762 Porto, Portugal\\
\email{mcunha@astro.up.pt, Mario.Monteiro@astro.up.pt}
\and 
Department of Physics and Astronomy, University of Aarhus, DK-8000 Aarhus C,
Denmark\\
\email{tct@phys.au.dk, hans@phys.au.dk, jcd@phys.au.dk, srf@phys.au.dk,
  karoff@phys.au.dk, \\fpp@phys.au.dk}
\and 
Sydney Institute for Astronomy (SIFA), School of Physics, University of Sydney 2006,
Australia\\
\email{bedding@Physics.usyd.edu.au}
\and
Laboratoire d'Astrophysique de Marseille, Traverse du Siphon BP8, 13376
Marseille Cedex 12, France\\
\email{bouchy@iap.fr}
\and 
Gemini Observatory, 670 N. A'ohoku Pl., Hilo, HI 96720,    USA\\
\email{tdall@eso.org}
\and 
Departamento de Matematica Aplicada da Faculdade de Ciencias da
Universidade do Porto, Portugal
\and 
Blackett Laboratory, Imperial College London, South Kensington, London
SW7 2BW, UK
             }

   \date{Received 5 August 2008; accepted 18 November 2008}

\newcommand{\half}{{\textstyle\frac{1}{2}}}
\newcommand{\Dnu}{\mbox{$\Delta \nu$}}
\newcommand{\GOLF}{{\em GOLF\/}}
\newcommand{\MOST}{{\em MOST\/}}
\newcommand{\SOHO}{{\em SOHO\/}}
\newcommand{\Teff}{\mbox{$T_{\rm eff}$}}
\newcommand{\VIRGO}{{\em VIRGO\/}}
\newcommand{\WIRE}{{\em WIRE\/}}
\newcommand{\acena}{\mbox{$\alpha$~Cen~A}}
\newcommand{\acenb}{\mbox{$\alpha$~Cen~B}}
\newcommand{\acen}{\mbox{$\alpha$~Cen}}
\newcommand{\bhyi}{\mbox{$\beta$~Hyi}}
\newcommand{\bvir}{\mbox{$\beta$~Vir}}
\newcommand{\cms}{\mbox{cm\,s$^{-1}$}}
\newcommand{\comment}[1]{{\bf [#1]}}
\newcommand{\dpav}{\mbox{$\delta$~Pav}}
\newcommand{\eboo}{\mbox{$\eta$~Boo}}
\newcommand{\ms}{\mbox{m\,s$^{-1}$}}
\newcommand{\muHz}{\mbox{$\mu$Hz}}
\newcommand{\muara}{\mbox{$\mu$~Ara}}
\newcommand{\mynote}[1]{{\bf\it [#1]}}
\newcommand{\nuind}{\mbox{$\nu$~Ind}}
\newcommand{\taucet}{\mbox{$\tau$~Cet}}
\newcommand{\new}[1]{{\bf #1}}
\renewcommand{\new}[1]{{\relax #1}}

  \abstract {We used HARPS to measure oscillations in the low-mass
star \taucet.  Although the data were compromised by instrumental noise, we
have been able to extract the main features of the oscillations.  We found
\taucet{} to oscillate with an amplitude that is about half that of the
Sun, and with a mode lifetime that is slightly shorter than solar.  The
large frequency separation is 169\,\muHz, and we have identified modes with
degrees 0, 1, 2, and~3.  We used the frequencies to estimate the mean
density of the star to an accuracy of 0.45\% which, combined with the
interferometric radius, gives a mass of $0.783 \pm 0.012\,M_{\sun}$ (1.6\%).}

   \keywords{Stars: oscillations -- Stars: individual: \taucet{}, \dpav, \acenb }

   \maketitle

\section{Introduction}

In the past few years, a new generation of high-resolution, high-precision
spectrographs has been providing unprecedented opportunities for studying
the fine details of stellar interiors and evolution through the detection
of tiny stellar oscillations. The observation and analysis of stellar
oscillations, or asteroseismology, has the potential to change dramatically
our views of stars.

The G8\,V star $\tau$ Ceti (HR\,509; HD\,10700; HIP\,8102; $V=3.50$) is
expected to have a convective envelope and therefore to display solar-like
oscillations.  Since \taucet{} has a lower metallicity than the Sun (${\rm
[Fe/H]} = -0.5 \pm 0.03$; \citealt{SKC98}), it bridges the gap towards very
metal-poor population II asteroseismic target stars such as $\nu$~Ind,
where solar-type oscillations have been detected \citep{BBC2006,CKB2007}.
Moreover, among stars for which a detection of solar-type oscillations have
been attempted (see \citealt{B+K2007c} and \citealt{AChDC2008} for recent
summaries), \taucet{} has the lowest mass.

\new{As a nearby bright star, \taucet\ has been intensively studied.  A
rotational period of 34 days is suggested by sporadic periodicities in
Ca~{\sc ii} \citep{BNRS96}, but overall \taucet\ is a very inactive star
with almost no rotational modulation.  This led \citet{G+B94} to propose
that \taucet\ is seen nearly pole-on, while \citet{JSC2004} have suggested
that it may be in a phase analogous to the solar Maunder minimum.  Its
stability makes it a favoured target for testing the velocity stability of
exoplanet programmes \citep[e.g.][]{BMW96}.  Despite many velocity
observations by different groups, no planetary companions have been
reported \citep{WEC2006}.  Direct imaging with the Hubble Space Telescope
also failed to detect a companion \citep{SGB2000}.  However,
\citet{GHW2004} have imaged a debris disc around \taucet\ that has a dust
mass at least an order of magnitude greater than in the Kuiper Belt.  }

We note that \taucet{} is particularly suitable for an asteroseismic
observing campaign because its radius has been determined
interferometrically with an accuracy of 0.5\%.  The combination of
interferometric and asteroseismic results has been applied to several other
stars,
as discussed in detail by \citet{CMM2007} and \citet{CAChD2007}.  As
stressed by \citet{B+G94}, for example, oscillation frequencies are most
valuable for testing evolution theories when the other fundamental stellar
properties are well-constrained.  \taucet{} satisfies this requirement as
well as can be done for any single star.

\begin{figure*}
\resizebox{\hsize}{!}{
\includegraphics{taucet-time-series.epsi}}
\caption{\label{fig.series} Time series of velocity measurements of
\taucet. }

\resizebox{\hsize}{!}{
\includegraphics{dpav-harps-time-series.epsi}}
\caption{\label{fig.dpav-series}  Velocity measurements of
\dpav{} over 1.5\,hr at the start of night~3. }
\end{figure*}

\section{Velocity observations and guiding noise}

We were allocated six nights to observe \taucet{} on 2004 October 2--7,
using the HARPS spectrograph (High Accuracy Radial velocity Planet
Searcher) on the 3.6-m telescope at the European Southern Observatory on La
Silla in Chile.  This spectrograph includes a thorium emission lamp to
provide a stable wavelength reference.  

We obtained 1962 spectra of \taucet, with a dead time of 31\,s between
exposures.  For the first two nights we used an exposure time of 40\,s
(resulting in a Nyquist frequency of \new{7.04\,mHz}) but shortened this to
23\,s (Nyquist frequency \new{9.26\,mHz}) for the remainder, in order to
sample better the noise at high frequencies (see below).  The velocities
were processed using the method described by \citet{BPQ2001} and the
resulting velocities are shown in Fig.~\ref{fig.series}.  The fourth and
fifth nights were mostly lost to poor weather.

For about 1.5\,hr at the start of each night, when \taucet{} was
inaccessible, we observed the star \dpav{} (HR\,7665; HD\,190248;
HIP\,99240; G6-8\,IV; $V=3.56$).  Small amounts of data on this star were
also obtained with UVES at the VLT and UCLES at the AAT by \citet{KBB2005},
who found oscillations centred at 2.3\,mHz with peak amplitudes close to
solar.  We obtained a total of 225 spectra of \dpav{} with HARPS, with
exposure times of 50\,s (nights~1 and~2) and 23\,s (nights~3 and~5).  The
velocities for night~3 (100 data points) are shown in
Fig.~\ref{fig.dpav-series}.

\begin{figure}[h!]
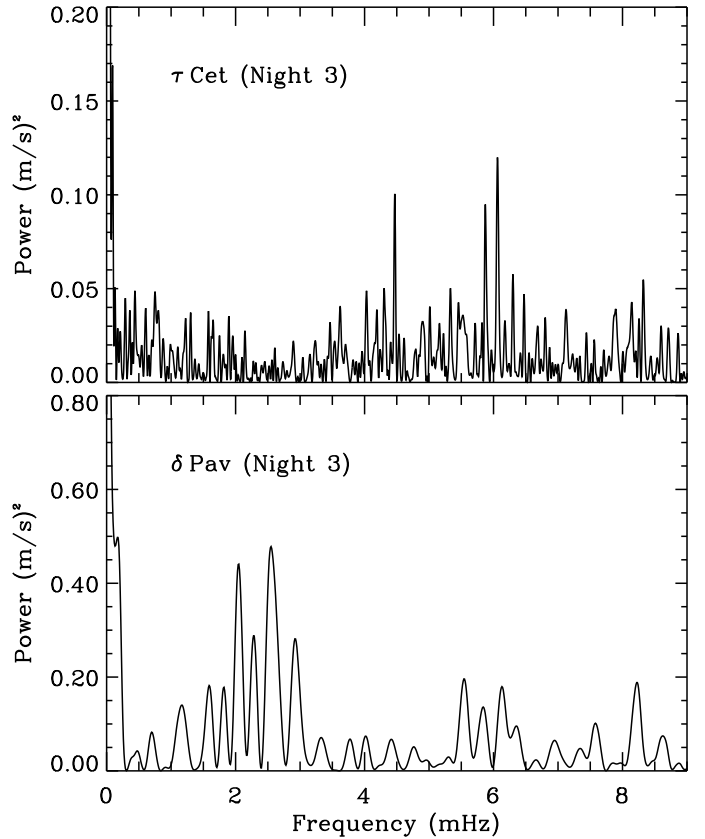

\resizebox{\hsize}{!}{\includegraphics{taucet-power-night3.epsi}}

\smallskip

\resizebox{\hsize}{!}{\includegraphics{dpav-harps-power.epsi}}
\caption{\label{fig.power-three} Power spectrum of velocity measurements
for night~3 only, for \taucet{} (9.1\,hr) and \dpav{} (1.5\,hr).   }
\end{figure}

{}From the scatter in the velocities and the noise in the power spectra for
both \taucet{} and \dpav, it was obvious that an unexpected noise source
was affecting the velocities.  Figure~\ref{fig.power-three} shows the power
spectrum for night~3 for both stars.  For \taucet{} (upper panel), there is
a clear excess at 4\,mHz, as expected for oscillations in this star.
However, there is also a significant power excess around 6\,mHz.  For
\dpav{} (lower panel of Fig.~\ref{fig.power-three}), the power centred at
about 2.2\,mHz is from oscillations (see \citealt{KBB2005}) but we again
see additional power at 6\,mHz.  The effect of this instrumental noise is
clearly visible by comparing the time series of \dpav{} in
Fig.~\ref{fig.dpav-series} of this paper with that in Fig.~2 of
\citet{KBB2005}.

\begin{figure*}[t]
\resizebox{\hsize}{!}{
\includegraphics{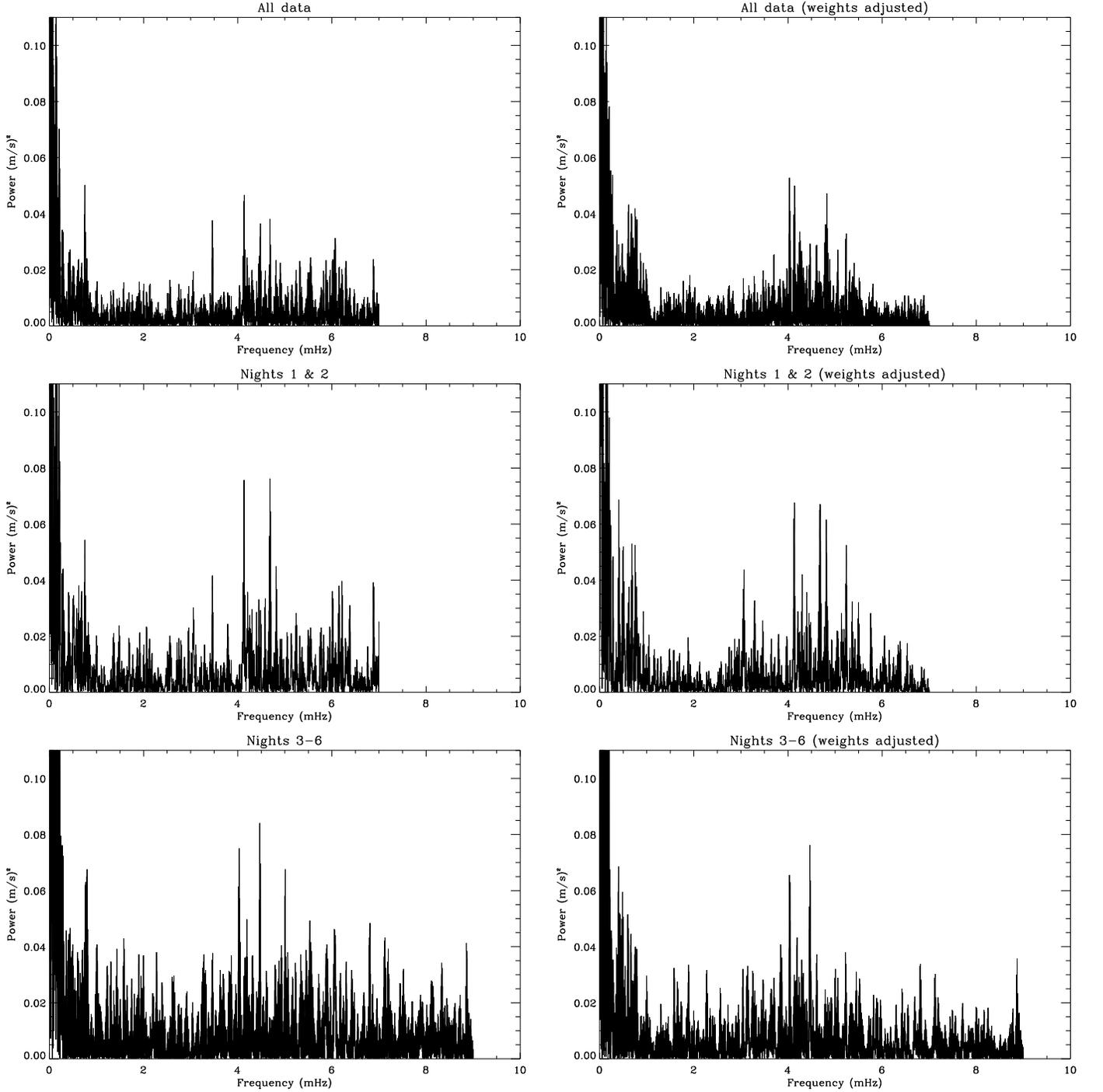}}
\caption{\label{fig.power} Power spectrum of \taucet{} for different
  subsets of the data.  The left panels show the spectrum with the use of
  the raw weights, and the right panels show the spectrum after the weights
  had been adjusted to account for bad data points and for night-to-night
  variations in the noise level (see text).}
\end{figure*}

The spurious signal at 6\,mHz was later traced to a periodic error in the
telescope guiding system.  \new{Noise spikes at 3.1 and 6.2\,mHz have been
reported} in HARPS observations of other oscillating stars, namely 70~Oph
\citep{C+E2006}, \acena{} \citep{BazBK2007} and \bhyi{} \citep{BKA2007}.
In the case of \taucet{}, the 6\,mHz noise is particularly problematic
because it covers a fairly broad range of frequencies and because the
stellar oscillations have very low amplitude. \new{There does not seem to
be a strong noise signal at 3\,mHz in our data, although it is difficult to
be certain.}


\section{Data analysis and results}

Our analysis of the velocity data and of the extracted power spectrum for
\taucet{} follows the method developed and used for \acena{}
\citep{BBK2004,BKB2004}, \acenb{} \citep{KBB2005}, \nuind{} \citep{BBC2006}
and \bhyi{} \citep{BKA2007}.  As usual, we have used the measurement
uncertainties, $\sigma_i$, as weights in calculating the power spectrum
(according to $w_i = 1/\sigma_i^2$).  The main difference between \taucet{}
and other stars that we have analysed, apart from the problem of excess
noise from the periodic guiding error, is the single-site nature of the
observations.  Because of this, we have not attempted to optimize the
weights to reduce the sidelobes, in the way that we did for other stars.

\subsection{Adjusting the weights}

As for previous stars, we adjusted the statistical weights to account for
bad data points and for night-to-night variations in the noise level.  We
did this by measuring the noise at frequencies where the oscillation signal
and the long-term drifts are negligible.  The first step was therefore to
remove all power below $800\,\mu$Hz (to avoid the slow drifts), as well as
all power between $3000\,\mu$Hz and $5500\,\mu$Hz (which is dominated by
the oscillations).  This filtering was performed by the standard method of
iterative sine-wave fitting (sometimes known as `pre-whitening').  In this
method, the highest peak is identified in the region of the power spectrum
that is to be removed, the corresponding sinusoid is subtracted from the
time series, the power spectrum is recomputed and the procedure is repeated
until all the power is removed.

Once this was done, the filtered time series for each night was examined
for bad data points.  These were identified as those deviating from the
mean scatter by more than 4-$\sigma$, and were re-assigned lower
statistical weights.  \new{We found that more than 10\% of data points had
to be significantly down-weighted.  This fraction is much greater than for
previous stars that we have observed and indicates the serious effects of
the guiding errors.}  At the same time, we scaled the statistical weights
on a night-to-night basis, in order to reflect the noise measured at high
frequencies.

The consequence of using the revised weights is a significant improvement
in the signal-to-noise, as can be seen in Fig.~\ref{fig.power}.  Comparison
of the left and right panels in that figure shows that the adjustment of
weights has removed essentially all the excess noise at 6--7\,mHz and also
decreased the noise level in the range 1--3\,mHz.  The mean noise in those
regions, measured in the amplitude spectrum of the whole data set, \new{was
reduced from 6.0 to} 4.0\,\cms.  \new{Most of this reduction came from the
down-weighting of bad data points, as described above.}  Also note that the
strongest oscillation peaks are not as high in the combined data (top panel
of Fig.~\ref{fig.power}) as in the shorter subsets (middle and bottom
panels).  This reflects the finite lifetime of the modes (see
Sec.~\ref{sec.freqs}).

\subsection{\label{powersec} The large separation}

The final power spectrum of \taucet{} is shown in \new{the top-right panel of}
Fig.~\ref{fig.power}.  There is a clear excess due to oscillations
which is centred at 4.5\,mHz.  The next step was to search for a regular
series of peaks, as expected for p-mode oscillations, and to measure the
large frequency separation,~$\Dnu$.  We did this in two ways.  The first
was to smooth the power spectrum and then calculate the autocorrelation in
the region of excess power, between 2.5 and 6.0\,mHz.  This produced a
clear peak at 169\,\muHz.

\begin{figure}
\resizebox{\hsize}{!}{
\includegraphics{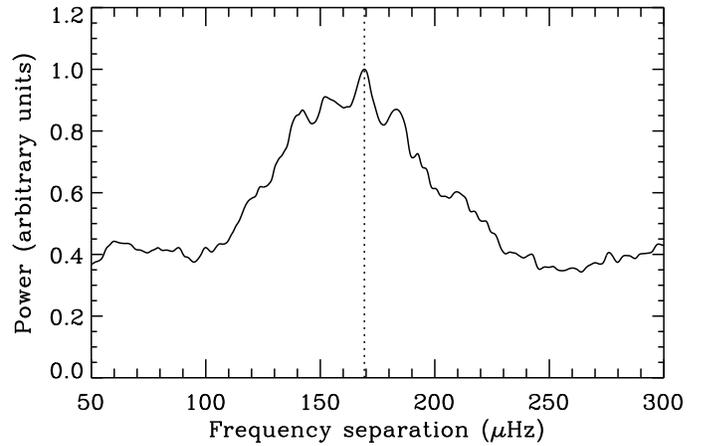}}
\caption{\label{fig.large-sep} Summed power of \taucet, using different
  values of the large separation.  The dotted line marks the maximum, which
  occurs at a frequency separation of 169.2\,\muHz.  }
\end{figure}

The second method, which we developed for the Kepler pipeline
\citep{ChDAB2007}, involved measuring the highest peak in the collapsed
power spectrum for a range of values of the large separation.  The
collapsed power spectrum for a given value of $\Dnu$ is calculated by
dividing the power spectrum into intervals of length $\Dnu$ and summing
these.  The result is shown in Fig.~\ref{fig.large-sep}, and again we see a
peak at 169\,\muHz. 

\subsection{Oscillations frequencies and mode lifetime} \label{sec.freqs}

\new{Mode frequencies for low-degree p-mode oscillations in main-sequence stars
are well approximated by a regular series of peaks, with frequencies given
by the following asymptotic relation:
\begin{equation}
  \nu_{n,l} = \Dnu (n + \half l + \epsilon) - l(l+1) D_0.
        \label{eq.asymptotic}
\end{equation}
Here $n$ (the radial order) and $l$ (the angular degree) are integers,
$\Dnu$ (the large separation) depends on the sound travel time across the
whole star, $D_0$ is sensitive to the sound speed near the core and
$\epsilon$ is sensitive to the surface layers.  See \citet{ChD2004} for a
recent review of the theory of solar-like oscillations.}

We have extracted the individual oscillation frequencies for \taucet{}
using the standard method of iterative sine-wave fitting \new{down to S/N =
2.5}.  The single-site nature of the observations and the relatively low
signal-to-noise ratio mean that this process is susceptible to
one-cycle-per-day ambiguities ($\pm 11.57\,\muHz$).  On the other hand, we
are helped greatly by the fact that both the large and small separations
are much greater than the expected mode linewidth, and so all modes are
well separated.  Furthermore, \taucet{} is an unevolved star and so we
expect the oscillation frequencies to follow quite closely the asymptotic
relation, without the presence of mixed modes.  We have used this
information to guide our selection of the correct peaks, but we stress that
there is some uncertainty in the correct mode identification.

\begin{figure}
\resizebox{\hsize}{!}{
\includegraphics{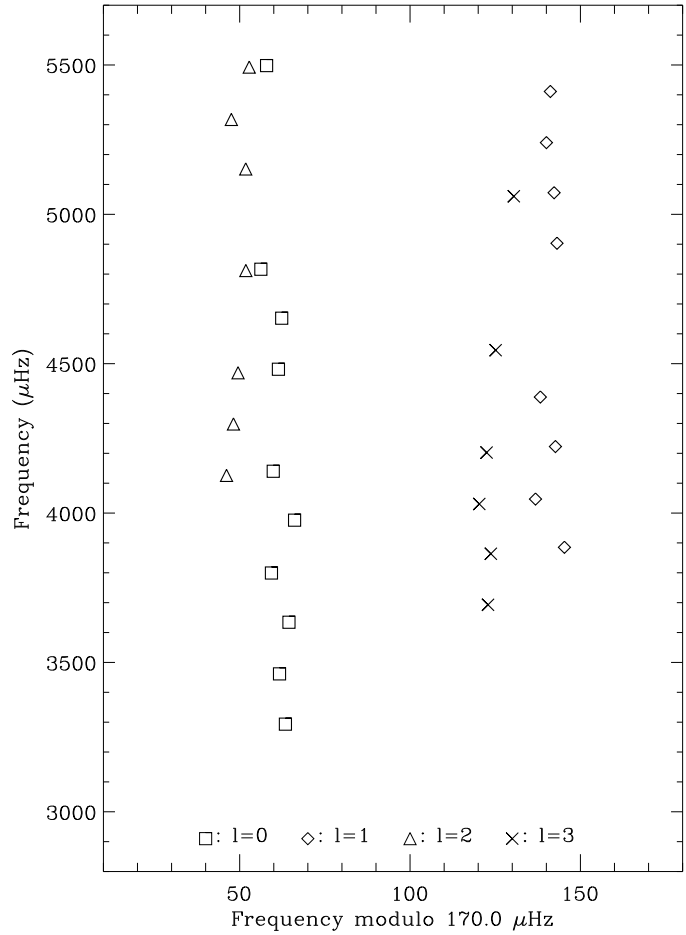}}
\caption{\label{fig.echelle} Echelle diagram of oscillation frequencies for
\taucet, \new{using the frequencies listed in Table~\ref{tab.freqs}.}  }
\end{figure}

The extracted frequencies are listed in Table~\ref{tab.freqs}.  They are
also shown in Fig.~\ref{fig.echelle} in echelle format, where the
frequencies are stacked in segments of length \Dnu.  As noted above, the
mode identification is uncertain, and this is particularly true for the
$l=0$ and $l=2$ modes above 5\,mHz.  

A fit to these frequencies provides the various large and small
separations, as listed in Table~\ref{tab.params}.  \new{For the definitions
of these separations see \citet{B+K2003}, for example.  The separations
generally vary with frequency and so the values in Table~\ref{tab.params} are
given at 4.3\,mHz.}  The scatter of these frequencies about smooth ridges
in the echelle diagram \new{is about 1--2\,\muHz, which indicates the
uncertainties in the individual frequencies in the table.}  From this
scatter we can estimate the mode lifetime, using the method described by
\citet{KBB2005}.  We find a value of $1.7\pm 0.5$\,d, which is slightly
lower than the value of $2.88\pm0.07$\,d measured for the Sun
\citep{CEI97}.

\begin{table}
\caption{\label{tab.freqs}Oscillation frequencies in \taucet\ \new{(in \muHz)}}
\centering
\begin{tabular}{c c c c c}
\hline\hline
$n$ & $l=0$ & $l=1$ & $l=2$ & $l=3$ \\
\hline
  18 &   3293.4 &          &          &          \\
  19 &   3461.7 &          &          &   3692.9 \\
  20 &   3634.5 &          &          &   3863.7 \\
  21 &   3799.3 &   3885.3 &          &   4030.3 \\
  22 &   3976.1 &   4046.8 &   4126.1 &   4202.5 \\
  23 &   4139.9 &   4222.7 &   4298.2 &          \\
  24 &          &   4388.3 &   4469.5 &   4545.1 \\
  25 &   4481.4 &          &          &          \\
  26 &   4652.3 &          &   4811.8 &          \\
  27 &   4816.2 &   4903.1 &          &   5060.5 \\
  28 &          &   5072.3 &   5151.8 &          \\
  29 &          &   5240.0 &   5317.5 &          \\
  30 &          &   5411.2 &   5492.8 &          \\
  31 &   5497.9 &          &          &          \\
\hline
\end{tabular}
\end{table}

\begin{table}
\caption{\label{tab.params} Oscillation parameters for \taucet{} \new{(see Sec.~\ref{sec.freqs})}}
\centering
\begin{tabular}{l r}
\hline\hline
Parameter & Value at 4.3\,mHz\\
\hline
$\Delta\nu_0$ (\muHz)          & $169.6 \pm 0.2$ \\
$\Delta\nu_1$ (\muHz)          & $170.0 \pm 0.3$ \\
$\Delta\nu_2$ (\muHz)          & $170.5 \pm 0.3$ \\
$\Delta\nu_3$ (\muHz)          & $171.0 \pm 0.3$ \\
$\delta\nu_{01}$ (\muHz)       &$ 4.7 \pm 1.3$   \\
$\delta\nu_{02}$ (\muHz)       &$12.7 \pm 1.2$   \\
$\delta\nu_{03}$ (\muHz)       &$21.4 \pm 1.2$   \\
$\delta\nu_{13}$ (\muHz)       &$16.7 \pm 1.4$   \\
$D_0$            (\muHz)       & $1.77 \pm 0.13$       \\
mode lifetime (d)              & $1.7\pm0.5$\\
\hline
\end{tabular}
\end{table}

\subsection{Stellar Parameters}

Detailed fitting of the oscillation frequencies of \taucet{} with
theoretical models is beyond the scope of this paper.  However, we can use
our results to determine the mean density of the star, via the empirical
method described by \citet{KBChD2008}.  \new{This method corrects the
frequencies of stellar models for near-surface effects by making use of the
fact that the offset between observations and models should tend to zero
with decreasing frequency.  The method involves fitting both \Dnu\ and the
absolute frequencies of the radial modes (i.e., those having degree $l=0$).
We applied the method to \taucet, using models computed with the Aarhus
stellar evolution code (ASTEC, \citealt{ChD2008a}) and the Aarhus adiabatic
oscillation package (ADIPLS, \citealt{ChD2008b}).  } The result is a mean
density for \taucet{} of $2.21 \pm 0.01$\,g\,cm$^{-3}$ (0.45\%).

As mentioned in the Introduction, the angular diameter of \taucet{} has
also been measured.  This was first done by \citet{PTG2003} with the VINCI
instrument on the VLTI\@.  They obtained an angular diameter, corrected for
limb darkening, of $1.97 \pm 0.05$\,mas (2.5\%), where the uncertainty was
dominated by the uncertainty in the angular diameter of the calibrator
star.  Subsequently, \citet{DiFTK2004} used the VLTI with smaller
calibrator stars to obtain an improved diameter of $2.032 \pm 0.031$\,mas
(1.5\%).  An even more accurate measurement was obtained by
\citet{DiFAA2007} using the FLUOR instrument on the CHARA array, giving
$2.015 \pm 0.011$\,mas (0.5\%).  The weighted mean of these measurements,
which we adopt here, is $2.022 \pm 0.010$\,mas (0.5\%).  Using the revised
{\em Hipparcos} parallax for \taucet{} of $273.96 \pm 0.170$\,mas
\citep{vanLee2007} gives a radius of $0.793 \pm 0.004\,R_{\sun}$ (0.5\%).
Finally, combining this radius with our estimate from asteroseismology of
the mean density gives a mass for \taucet{} of $0.783 \pm 0.012\,M_{\sun}$
(1.6\%).

For convenience, we also give an estimate of the luminosity of \taucet.
The apparent visual magnitude of $V=3.50\pm0.01$, with the revised parallax, gives
an absolute magnitude of $M_V = 5.69\pm0.01$.  Using a bolometric correction for
\taucet{} of $-0.17 \pm 0.02$ \citep{CPF2006} and adopting an absolute
bolometric magnitude for the Sun of $M_{\rm bol, \sun} = 4.74$
\citep{BCP98}, we derive a luminosity for \taucet{} of $0.488 \pm
0.010\,L_{\sun}$ (2.0\%).

\subsection{Oscillation amplitudes}  \label{sec.amp}

\begin{figure}
\resizebox{\hsize}{!}{
\includegraphics{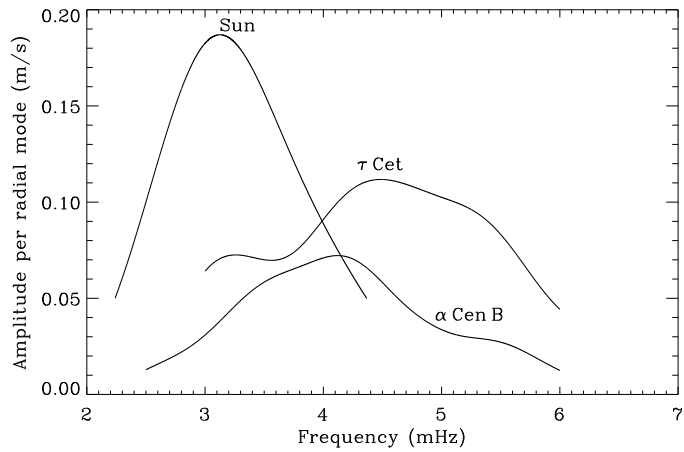}}
\caption{\label{fig.amp} Smoothed amplitude curve for \taucet.  For
  comparison, similar curves are shown for the Sun and \acenb{}
  \citep{KBA2008}.}
\end{figure}

We have determined the oscillation amplitude per mode from the smoothed
power spectrum, using the method described by \citet{KBA2008}.  This
produces a result that is independent of the stochastic nature of the
excitation and damping.  The result is shown in Fig.~\ref{fig.amp}.  The
peak of the envelope occurs at frequency $\nu_{\rm max} = 4.49$\,mHz and
the peak amplitude per mode (for radial oscillations) is $11.2 \pm
0.8$\,cm/s, where the uncertainty is estimated using Eq.~(3) of
\citet{KBA2008}.  For comparison, we also show in Fig.~\ref{fig.amp} the
amplitude curves measured for the Sun (using stellar techniques) and for
\acenb, both taken from Fig.~8 of \citet{KBA2008}.  Note that the
luminosity and mass of \acenb{} are, respectively, $0.51\,L_{\sun}$
\citep{Yil2007} and $0.93\,M_{\sun}$ \citep{PNMcC2002}.  With this in mind,
we see that the relative amplitudes of the three stars in
Fig.~\ref{fig.amp} are in reasonable agreement with the $L/M$ scaling
relation proposed by \citet[][see also \citealt{SGT2007}]{K+B95}

\section{Conclusions}

We have used HARPS to measure oscillations in the low-mass star \taucet.
Although the data were compromised by instrumental noise, we have been able
to extract the main features of the oscillations.  We found \taucet{} to
oscillate with an amplitude that is about half that of the Sun, and with a
mode lifetime that is slightly smaller than solar.  The large frequency
separation is 169\,\muHz, and we have identified modes with degrees 0, 1, 2
and~3.  It is important to stress that, given the relatively low
signal-to-noise ratio and the single-site nature of the observations, there
is some uncertainty in the correct mode identification.

We used the frequencies of the radial modes to estimate the mean
density of the star to an accuracy of 0.45\%, from which we derived a mass
of $0.783 \pm 0.012\,M_{\sun}$ (1.6\%).  More detailed modelling of the
oscillation frequencies will be the subject of a future paper.

\begin{acknowledgements}
We thank the referee for helpful suggestions.
This work has been supported by the Danish Natural Science Research Council
and the Australian Research Council.  MC is supported by the Ciencia2007
Programme from FCT (C2007-CAUP-FCT/136/2006).
\end{acknowledgements}



\begin{thebibliography}{38}
\expandafter\ifx\csname natexlab\endcsname\relax\def\natexlab#1{#1}\fi

\bibitem[{{Aerts} {et~al.}(2008){Aerts}, {Christensen-Dalsgaard}, {Cunha}, \&
  {Kurtz}}]{AChDC2008}
{Aerts}, C., {Christensen-Dalsgaard}, J., {Cunha}, M., \& {Kurtz}, D.~W. 2008,
  Sol. Phys., 251, 3

\bibitem[{{Baliunas} {et~al.}(1996){Baliunas}, {Nesme-Ribes}, {Sokoloff}, \&
  {Soon}}]{BNRS96}
{Baliunas}, S.~L., {Nesme-Ribes}, E., {Sokoloff}, D., \& {Soon}, W.~H. 1996,
  ApJ, 460, 848

\bibitem[{{Bazot} {et~al.}(2007){Bazot}, {Bouchy}, Kjeldsen, Charpinet,
  Laymand, \& {Vauclair}}]{BazBK2007}
{Bazot}, M., {Bouchy}, F., Kjeldsen, H., {et~al.} 2007, A\&A, 470, 295

\bibitem[{Bedding \& Kjeldsen(2003)}]{B+K2003}
Bedding, T.~R. \& Kjeldsen, H. 2003, PASA, 20, 203

\bibitem[{{Bedding} \& {Kjeldsen}(2007)}]{B+K2007c}
{Bedding}, T.~R. \& {Kjeldsen}, H. 2007, in Unsolved Problems in Stellar
  Physics: A Conference in Honour of Douglas Gough, AIP Conf. Proc., vol. 948,
  ed. R.~J. Stancliffe, G.~Houdek, R.~G. Martin, \& C.~A. Tout, 117

\bibitem[{Bedding {et~al.}(2004)Bedding, Kjeldsen, Butler, {et~al.}}]{BKB2004}
Bedding, T.~R., Kjeldsen, H., Butler, R.~P., {et~al.} 2004, ApJ, 614, 380

\bibitem[{{Bedding} {et~al.}(2006){Bedding}, {Butler}, {Carrier},
  {et~al.}}]{BBC2006}
{Bedding}, T.~R., {Butler}, R.~P., {Carrier}, F., {et~al.} 2006, ApJ, 647, 558

\bibitem[{{Bedding} {et~al.}(2007){Bedding}, {Kjeldsen}, Arentoft,
  {et~al.}}]{BKA2007}
{Bedding}, T.~R., {Kjeldsen}, H., Arentoft, T., {et~al.} 2007, ApJ, 663, 1315

\bibitem[{Bessell {et~al.}(1998)Bessell, {Castelli}, \& {Plez}}]{BCP98}
Bessell, M.~S., {Castelli}, F., \& {Plez}, B. 1998, A\&A, 333, 231, erratum:
  337, 321

\bibitem[{{Bouchy} {et~al.}(2001){Bouchy}, {Pepe}, \& {Queloz}}]{BPQ2001}
{Bouchy}, F., {Pepe}, F., \& {Queloz}, D. 2001, A\&A, 374, 733

\bibitem[{Brown \& Gilliland(1994)}]{B+G94}
Brown, T.~M. \& Gilliland, R.~L. 1994, ARA\&A, 32, 37

\bibitem[{Butler {et~al.}(1996)Butler, Marcy, Williams, McCarthy, Dosanjh, \&
  Vogt}]{BMW96}
Butler, R.~P., Marcy, G.~W., Williams, E., {et~al.} 1996, PASP, 108, 500

\bibitem[{Butler {et~al.}(2004)Butler, Bedding, Kjeldsen, {et~al.}}]{BBK2004}
Butler, R.~P., Bedding, T.~R., Kjeldsen, H., {et~al.} 2004, ApJ, 600, L75

\bibitem[{{Carrier} \& {Eggenberger}(2006)}]{C+E2006}
{Carrier}, F. \& {Eggenberger}, P. 2006, A\&A, 450, 695

\bibitem[{{Carrier} {et~al.}(2007){Carrier}, {Kjeldsen}, {Bedding},
  {et~al.}}]{CKB2007}
{Carrier}, F., {Kjeldsen}, H., {Bedding}, T.~R., {et~al.} 2007, A\&A, 470, 1059

\bibitem[{{Casagrande} {et~al.}(2006){Casagrande}, {Portinari}, \&
  {Flynn}}]{CPF2006}
{Casagrande}, L., {Portinari}, L., \& {Flynn}, C. 2006, MNRAS, 373, 13

\bibitem[{Chaplin {et~al.}(1997)Chaplin, {Elsworth}, {Isaak}, {McLeod},
  {Miller}, \& {New}}]{CEI97}
Chaplin, W.~J., {Elsworth}, Y., {Isaak}, G.~R., {et~al.} 1997, MNRAS, 288, 623

\bibitem[{Christensen-Dalsgaard(2004)}]{ChD2004}
Christensen-Dalsgaard, J. 2004, Sol. Phys., 220, 137

\bibitem[{Christensen-Dalsgaard(2008a)}]{ChD2008a}
Christensen-Dalsgaard, J. 2008a, Ap\&SS, 316, 13

\bibitem[{{Christensen-Dalsgaard}(2008b)}]{ChD2008b}
{Christensen-Dalsgaard}, J. 2008b, Ap\&SS, 316, 113

\bibitem[{{Christensen-Dalsgaard} {et~al.}(2007){Christensen-Dalsgaard},
  {Arentoft}, {Brown}, {Gilliland}, {Kjeldsen}, {Borucki}, \&
  {Koch}}]{ChDAB2007}
{Christensen-Dalsgaard}, J., {Arentoft}, T., {Brown}, T.~M., {et~al.} 2007,
  Commun. Asteroseismology, 150, 350

\bibitem[{{Creevey} {et~al.}(2007){Creevey}, {Monteiro}, {Metcalfe},
  {et~al.}}]{CMM2007}
{Creevey}, O.~L., {Monteiro}, M.~J.~P.~F.~G., {Metcalfe}, T.~S., {et~al.} 2007,
  ApJ, 659, 616

\bibitem[{{Cunha} {et~al.}(2007){Cunha}, {Aerts}, {Christensen-Dalsgaard},
  {et~al.}}]{CAChD2007}
{Cunha}, M.~S., {Aerts}, C., {Christensen-Dalsgaard}, J., {et~al.} 2007, A\&AR,
  14, 217

\bibitem[{{Di Folco} {et~al.}(2004){Di Folco}, {Th{\'e}venin}, {Kervella},
  {Domiciano de Souza}, {Coud{\'e} du Foresto}, {S{\'e}gransan}, \&
  {Morel}}]{DiFTK2004}
{Di Folco}, E., {Th{\'e}venin}, F., {Kervella}, P., {et~al.} 2004, A\&A, 426,
  601

\bibitem[{{Di Folco} {et~al.}(2007){Di Folco}, {Absil}, {Augereau},
  {M{\'e}rand}, {Coud{\'e} Du Foresto}, {et~al.}}]{DiFAA2007}
{Di Folco}, E., {Absil}, O., {Augereau}, J.-C., {et~al.} 2007, A\&A, 475, 243

\bibitem[{{Gray} \& {Baliunas}(1994)}]{G+B94}
{Gray}, D.~F. \& {Baliunas}, S.~L. 1994, ApJ, 427, 1042

\bibitem[{{Greaves} {et~al.}(2004){Greaves}, {Wyatt}, {Holland}, \&
  {Dent}}]{GHW2004}
{Greaves}, J.~S., {Wyatt}, M.~C., {Holland}, W.~S., \& {Dent}, W.~R.~F. 2004,
  MNRAS, 351, L54

\bibitem[{{Judge} {et~al.}(2004){Judge}, {Saar}, {Carlsson}, \&
  {Ayres}}]{JSC2004}
{Judge}, P.~G., {Saar}, S.~H., {Carlsson}, M., \& {Ayres}, T.~R. 2004, ApJ,
  609, 392

\bibitem[{Kjeldsen \& Bedding(1995)}]{K+B95}
Kjeldsen, H. \& Bedding, T.~R. 1995, A\&A, 293, 87

\bibitem[{Kjeldsen {et~al.}(2005)Kjeldsen, Bedding, Butler, {et~al.}}]{KBB2005}
Kjeldsen, H., Bedding, T.~R., Butler, R.~P., {et~al.} 2005, ApJ, 635, 1281

\bibitem[{Kjeldsen {et~al.}(2008{\natexlab{a}})Kjeldsen, {Bedding}, Arentoft,
  {et~al.}}]{KBA2008}
Kjeldsen, H., {Bedding}, T.~R., Arentoft, T., {et~al.} 2008{\natexlab{a}}, ApJ,
  682, 1370

\bibitem[{Kjeldsen {et~al.}(2008{\natexlab{b}})Kjeldsen, Bedding, \&
  Christensen-Dalsgaard}]{KBChD2008}
Kjeldsen, H., Bedding, T.~R., \& Christensen-Dalsgaard, J. 2008{\natexlab{b}},
  ApJ, 683, L175

\bibitem[{Pijpers {et~al.}(2003)Pijpers, {Teixeira}, {Garcia}, {Cunha},
  {Monteiro}, \& {Christensen-Dalsgaard}}]{PTG2003}
Pijpers, F.~P., {Teixeira}, T.~C., {Garcia}, P.~J., {et~al.} 2003, A\&A, 406,
  L15

\bibitem[{{Pourbaix} {et~al.}(2002){Pourbaix}, {Nidever}, {McCarthy}, {Butler},
  {Tinney}, {Marcy}, {Jones}, {Penny}, {Carter}, {Bouchy}, {Pepe}, {Hearnshaw},
  {Skuljan}, {Ramm}, \& {Kent}}]{PNMcC2002}
{Pourbaix}, D., {Nidever}, D., {McCarthy}, C., {et~al.} 2002, A\&A, 386, 280

\bibitem[{{Samadi} {et~al.}(2007){Samadi}, {Georgobiani}, {Trampedach R.},
  {Goupil}, {Stein}, \& {Nordlund}}]{SGT2007}
{Samadi}, R., {Georgobiani}, D., {Trampedach R.}, R., {et~al.} 2007, A\&A, 463,
  297

\bibitem[{{Schroeder} {et~al.}(2000){Schroeder}, {Golimowski}, {Brukardt},
  {Burrows}, {Caldwell}, {Fastie}, {Ford}, {Hesman}, {Kletskin}, {Krist},
  {Royle}, \& {Zubrowski}}]{SGB2000}
{Schroeder}, D.~J., {Golimowski}, D.~A., {Brukardt}, R.~A., {et~al.} 2000, AJ,
  119, 906

\bibitem[{{Soubiran} {et~al.}(1998){Soubiran}, {Katz}, \& {Cayrel}}]{SKC98}
{Soubiran}, C., {Katz}, D., \& {Cayrel}, R. 1998, A\&AS, 133, 221

\bibitem[{van Leeuwen(2007)}]{vanLee2007}
van Leeuwen, F. 2007, {Hipparcos, the New Reduction of the Raw Data} (Springer:
  Dordrecht)

\bibitem[{{Wittenmyer} {et~al.}(2006){Wittenmyer}, {Endl}, {Cochran}, {Hatzes},
  {Walker}, {Yang}, \& {Paulson}}]{WEC2006}
{Wittenmyer}, R.~A., {Endl}, M., {Cochran}, W.~D., {et~al.} 2006, AJ, 132, 177

\bibitem[{{Y{\i}ld{\i}z}(2007)}]{Yil2007}
{Y{\i}ld{\i}z}, M. 2007, MNRAS, 374, 1264

\end{thebibliography}
\end{document}